%% file: main.tex
\documentclass[10pt]{article} % For LaTeX2e
\usepackage[preprint]{tmlr}
\input{math_commands.tex}

\usepackage{hyperref}
\usepackage{url}
\usepackage{graphicx}
\usepackage[margin=1in]{geometry}
\usepackage[utf8]{inputenc}
\usepackage{amsmath,amssymb,amsfonts}
\usepackage{graphicx}
\usepackage{xcolor}
\hypersetup{
    colorlinks  = true,
    linkcolor   = blue,
    citecolor   = blue,
    urlcolor    = blue
}
\usepackage{float}
\usepackage{titlesec}
\usepackage{tabularx}
\title{Beyond the Black Box: Interpretability of LLMs in Finance\thanks{The views expressed in this paper are those of the authors and do not necessarily reflect the views of Barclays.}}

\author{
  \begin{minipage}[t]{0.45\textwidth}
    \centering
    \textbf{Hariom Tatsat} \\
    \small hariom.x.tatsat@barclays.com \\
    \small Quantitative Analytics, Barclays
  \end{minipage}
  \hfill
  \begin{minipage}[t]{0.45\textwidth}
    \centering
    \textbf{Ariye Shater} \\
    \small ariye.shater@barclays.com \\
    \small Quantitative Analytics, Barclays
  \end{minipage}
}

\def\month{MM}  % Insert correct month for camera-ready version
\def\year{YYYY} % Insert correct year for camera-ready version
\def\openreview{\url{https://openreview.net/forum?id=XXXX}} % Insert correct link to OpenReview for camera-ready version

\begin{document}

\maketitle

\begin{abstract}

Large Language Models (LLMs) exhibit remarkable capabilities across a spectrum of tasks in financial services, including report generation, chatbots, sentiment analysis, regulatory compliance, algorithmic trading enhancement, investment advisory, financial knowledge retrieval, and summarization. However, their intrinsic complexity and lack of transparency pose significant challenges, especially in the highly regulated financial sector, where interpretability, fairness, and accountability are critical. 

As far as we are aware, this paper presents the \textbf{first application in the finance domain} of understanding and utilizing the inner workings of LLMs through \textbf{mechanistic interpretability}, addressing the pressing need for transparency and control in AI systems. Mechanistic interpretability is the most intuitive and transparent way to understand LLM behavior by reverse-engineering their internal workings. By dissecting the activations and circuits within these models, it provides insights into how specific features or components influence predictions—making it possible not only to observe but also to modify model behavior.

In this paper, we explore the theoretical aspects of mechanistic interpretability and demonstrate its practical relevance through a range of financial use cases and experiments, including applications in \textbf{trading strategies, sentiment analysis, bias, and hallucination detection.}

While not yet widely adopted, mechanistic interpretability is expected to become increasingly vital as adoption of LLMs increases. Advanced interpretability tools can ensure AI
systems remain ethical, transparent, and aligned with evolving financial regulations. In this
paper, we have put special emphasis on how these techniques can help unlock interpretability requirements for regulatory and compliance purposes—addressing both current needs and anticipating future expectations from financial regulators globally.

\end{abstract}

\section{Introduction}

As large language models (LLMs) and other AI systems become more sophisticated and deeply embedded in financial decision-making, ensuring their transparency, reliability, and alignment with regulatory and ethical standards is paramount. Financial institutions deploy AI models including LLMs across risk management, fraud detection, credit scoring, regulatory compliance, and algorithmic trading, often relying on complex black-box models \citep{Cerneviciene2024}. The lack of interpretability in these models poses significant risks, including bias, hallucinations, non-compliant decisions, and opaque reasoning, which can lead to regulatory penalties and financial losses \citep{Guidotti2018}. These concerns have driven the increasing adoption of Explainable AI (XAI) methodologies, with a significant shift toward mechanistic interpretability---a discipline that aims to reverse-engineer AI models and LLMs into understandable components rather than relying on post hoc explanations \citep{Rauker2023}.

Historically, interpretability in finance relied on feature importance methods, Shapley values (SHAP), and rule-based explainability frameworks \citep{Linardatos2020}. While these methods provided insights into model behavior, they lacked the ability to unravel the underlying computations and internal representations governing AI-driven financial decisions. Recent breakthroughs in mechanistic interpretability have enabled researchers to dissect large AI systems, such as transformers and deep neural networks, at a granular level, exposing how models understand several concepts \citep{Anthropic2024}. Notably, OpenAI's discovery of a ``sentiment neuron'' within an unsupervised language model exemplifies how specific neurons can encapsulate complex features like sentiment, highlighting the potential for mechanistic interpretability to reveal such structures \citep{Radford2017}.

Mechanistic interpretability provides two primary approaches: \emph{observation} and \emph{intervention}. Observation techniques, such as logit attribution and feature retrieval, enable researchers to identify which model components contribute to a given prediction. Intervention techniques, such as activation patching and feature steering via sparse autoencoders, allow targeted modifications to mitigate bias, reduce hallucinations, and enhance regulatory compliance. Sparse autoencoders, in particular, play a transformative role by extracting sparse feature representations, which not only enable the retrieval of relevant financial signals but also facilitate the suppression of unwanted or non-compliant outputs \citep{Bereska2024}. A related approach is dictionary learning, as discussed by \citet{hinton2023monosemantic}, where models learn to represent data using a set of basis functions or atoms, enabling more efficient representation and retrieval of financial signals. This approach can be particularly useful in financial models that require interpretable feature extraction, as it focuses on identifying distinct, monosemantic features that can explain complex predictions.

This shift toward mechanistic analysis is particularly critical for high-stakes financial applications. Traditional black-box approaches often require extensive prompt engineering or model fine-tuning to align AI outputs with compliance standards and risk frameworks. Mechanistic interpretability, by contrast, enables dynamic output adjustments while preserving transparency, significantly reducing the need for manual intervention. We demonstrate the impact of these techniques across key financial applications.

This paper is structured as follows: \textbf{Section 2} provides a detailed background on mechanistic interpretability, including foundational concepts and prior interpretability approaches in finance. \textbf{Section 3} outlines core interpretability techniques—logit attribution, activation patching, and sparse autoencoder-based techniques - applying both observation and intervention methodologies. \textbf{Section 4} introduces how mechanistic interpretability concepts can be applied in finance, while \textbf{Sections 5 and 6} demonstrate the results of our research across real-world applications including credit risk analysis, bias detection, and hallucination control.

\section{Background}

\subsection{Interpretability In General }
In order to promote the integration of AI into critical decision-making processes, interpretability and transparency have emerged as key priorities. Explainable AI (XAI) aims to improve trust and accountability by enabling users to understand, validate, and control AI models \citep{Adadi2018, Arrieta2020}. Early rule-based expert systems provided inherent interpretability, but the rise of deep learning (DL) models introduced a trade-off between performance and explainability \citep{Saeed2023}.  

Black-box models, such as deep neural networks (DNNs) deliver state-of-the-art accuracy but lack transparency, raising concerns in high-risk applications like finance and healthcare \citep{Linardatos2020}. In contrast, white-box models, including linear models, decision trees, and rule-based algorithms, provide structured and interpretable outcomes but may lack predictive power and fail to capture intricate feature dependencies.  

XAI techniques are categorized by their scope (local vs. global), timing (ante-hoc vs. post-hoc), and dependency (model-specific vs. model-agnostic). Ante-hoc methods integrate interpretability during training, while post-hoc approaches explain pre-trained models through techniques like SHAP (Shapley Additive Explanations), LIME (Local Interpretable Model-Agnostic Explanations), and attention visualization \citep{Saranya2023}. Model-agnostic methods generalize across architectures, whereas model-specific techniques leverage internal mechanisms to enhance transparency \citep{Zolanvari2021}.  

Regulatory frameworks such as the European Union's AI Act emphasize the need for fairness, transparency, and accountability in AI-driven decision-making \href{https://digital-strategy.ec.europa.eu/en/policies/european-approach-artificial-intelligence} {\citep{EU2020}}. However, achieving explainability remains a challenge, as increasing model complexity often leads to reduced interpretability. Researchers continue to explore ways to balance explainability, performance, and computational efficiency, ensuring AI remains both powerful and accountable \citep{Frasca2024}.  

\subsection{Interpretability Categories}

\begin{figure}[H]
    \centering
    \includegraphics[width=0.8\textwidth]{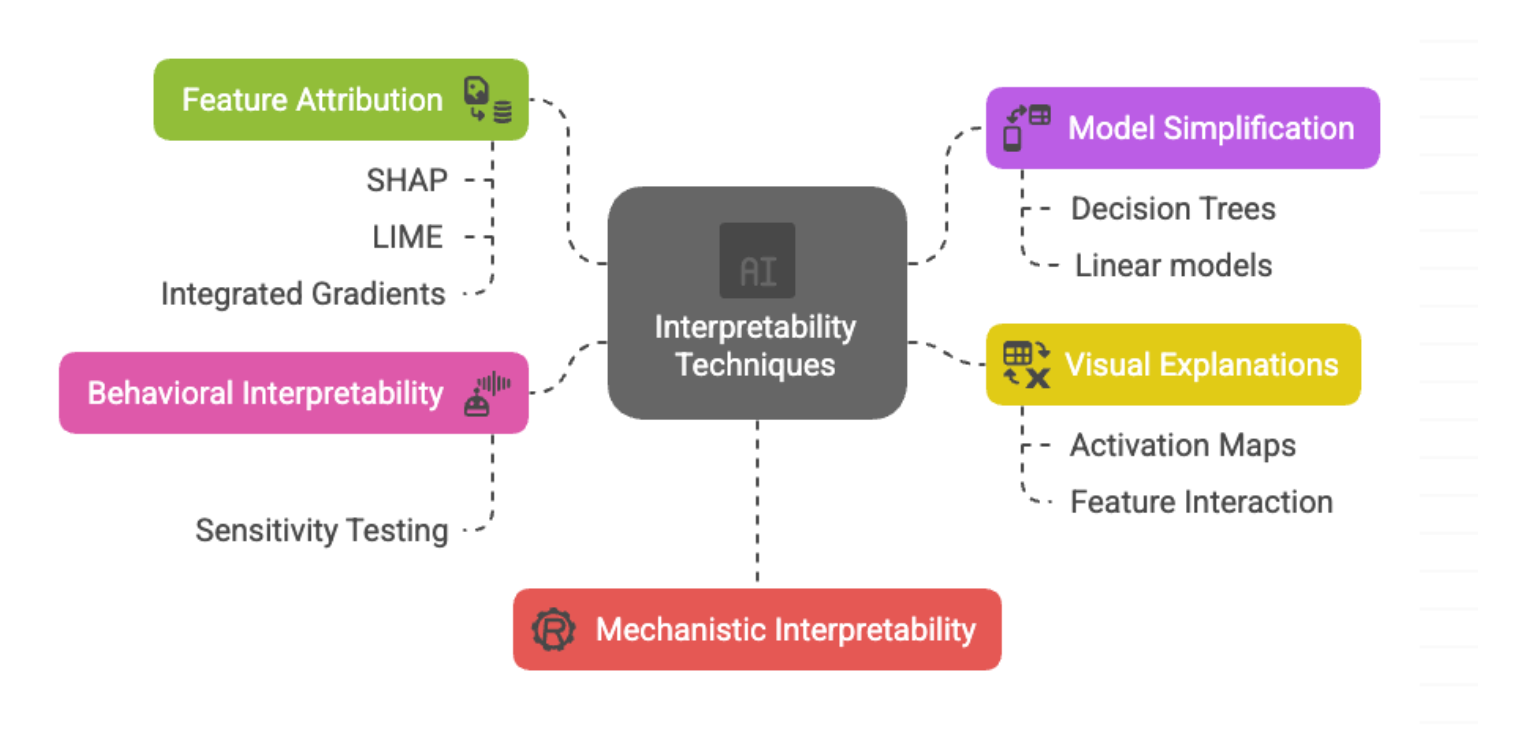}
    \caption{Interpretability techniques}
    \label{fig:interpretability_taxonomy}
\end{figure}

We encounter a spectrum of interpretability paradigms for decoding AI systems’ decision-making, ranging from external black-box techniques to internal analysis as shown in Figure \ref{fig:interpretability_taxonomy}. We contrast these paradigms with mechanistic interpretability, highlighting its distinct causal bottom-up perspective within the broader interpretability landscape. 
\subsubsection*{Feature Attribution Based Interpretability}
Feature attribution methods evaluate the significance of each input feature in influencing a model's predictions, helping understand decision boundaries and model behavior. Shapley Additive Explanations (SHAP) \citep{Lundberg2017} allocate feature contributions using cooperative game theory, while Local Interpretable Model-Agnostic Explanations (LIME) \citep{Ribeiro2016} approximate complex models with locally interpretable linear models. Gradient-based techniques such as Integrated Gradients, DeepLIFT, and GradCAM \citep{Sundararajan2017} quantify how changes in input affect outputs. These methods enhance interpretability by identifying key decision-making factors, though they often fail to capture deeper causal structures within neural networks.

\subsubsection*{Model Simplification-Based Interpretability} Model simplification techniques aim to retain predictive power while improving transparency by approximating black-box models with interpretable alternatives. Decision trees and rule-based models \citep{Hayashi2016} generate structured, human-readable rules for applications such as credit risk assessment. Knowledge distillation \citep{Zhang2020} compresses knowledge from complex models into simpler ones, preserving accuracy while improving interpretability. Although effective, these methods often fail to capture intricate feature dependencies present in high-dimensional financial data.

\subsubsection*{Visual Explanations-Based Interpretability}

Visual Explanations aim to create visual representations that simplify understanding how a model works, even when dealing with complex dimensions. These methods highlight the interaction
among features, identify the regions of an input that influence model outputs and use activation maps \citep{Selvaraju2016} to aid transparency.

\subsubsection*{Behavioral Interpretability}

Behavioral interpretability treats AI models as black boxes, analyzing how outputs change in response to different inputs. Techniques such as minimal pair analysis \citep{Warstadt2020} compare model responses to slight input modifications, while perturbation and sensitivity testing \citep{Casalicchio2018} assess model robustness to distributional shifts. Although useful for identifying biases and performance vulnerabilities, these methods provide limited insights into internal model structures.

\subsubsection*{Mechanistic Interpretability}

Mechanistic interpretability seeks to reverse-engineer AI models by analyzing circuits, neurons, and computational pathways, making it distinct from post-hoc explanations. \emph{Circuit-level analysis} \citep{elhage2021mathematical, nanda2023progress} identifies how specific patterns of attention heads and MLP neurons implement distinct subfunctions of a model, such as induction heads or induction circuits. \emph{Sparse autoencoders} \citep{Bereska2024} extract interpretable and disentangled features from model activations, enabling applications such as bias correction, steering, and concept-based debugging. \emph{Activation patching} \citep{elhage2021mathematical} allows interventions at intermediate layers of computation, letting researchers test hypotheses about how information flows and which components are causally responsible for outputs. This paradigm is especially critical in high-stakes domains such as finance and policy, where interpretability must go beyond surface-level explanations to ensure trust and accountability.

\subsection{Interpretability In Finance}

Financial AI models are integral to modern decision-making, driving risk management, fraud detection, credit assessments, and algorithmic trading and many other financial applications. However, their complexity raises concerns about transparency, compliance, and accountability. This section explores how interpretability techniques are applied in finance, aligning them with the key categories of explainability: feature attribution, model simplification, visual and concept-based methods, behavioral analysis, and mechanistic interpretability. By mapping explainability techniques to financial tasks, we illustrate how interpretability ensures regulatory compliance, enhances transparency, and mitigates financial risks.

\subsubsection*{Feature Attribution}

Feature attribution techniques assess the influence of individual financial attributes on model predictions, ensuring fairness and transparency. For example, models like SHAP \citep{Lundberg2017} and LIME \citep{Ribeiro2016} quantify the impact of financial variables such as income, debt levels, and credit history in determining loan approvals. Additionally, SHAP values are used in fraud detection models \citep{Zhang2020} to analyze the impact of transaction patterns on fraud risk. Gradient-based attribution techniques like Integrated Gradients \citep{Sundararajan2017} highlight critical risk factors affecting financial forecasts, though they do not explain deeper causal relationships.

\subsubsection*{Model Simplification}

Model simplification methods approximate complex AI models with interpretable surrogates, balancing accuracy and transparency. In regulatory compliance, decision trees and rule-based models \citep{Hayashi2016} offer structured rules for credit risk assessment, bankruptcy prediction, and fraud detection. Additionally, techniques like distillation \citep{Zhang2020} compress deep learning models into interpretable structures, making AI-driven financial systems more transparent while retaining predictive accuracy.

\subsubsection*{Visual Interpretability in Algorithmic Trading}

Visual interpretability helps investors and analysts understand AI-driven market predictions. For example, attention mechanisms \citep{Selvaraju2016} highlight key market indicators affecting stock movements.

\subsubsection*{Behavioral Interpretability For Market Stability And Fraud Detection}

Behavioral interpretability treats AI models as black boxes, examining how they react to input changes. Sensitivity and perturbation analysis \citep{Casalicchio2018} evaluates how AI-driven financial strategies perform under market shifts, while minimal pair analysis \citep{Warstadt2020} helps identify inconsistencies and biases in AI-driven fraud detection systems.

As shown in Table \ref{tab:financial_use_cases}, the financial applications are mapped to various interpretability techniques, including feature attribution, model simplification, and visual/concept-based approaches as well as behavioral analysis.

\begin{table}[h!]
\centering
\resizebox{\textwidth}{!}{%
\begin{tabular}{|l|l|l|l|l|}
\hline
\textbf{\large Financial Application} & \textbf{\large Feature Attribution} & \textbf{\large Model Simplification} & \textbf{\large Visual/Concept-Based} & \textbf{\large Behavioral} \\ \hline
\textbf{Credit Scoring}      & SHAP, LIME                   & Decision Trees, Rule-Based   & Attention Mechanisms          & Counterfactual Analysis      \\ \hline
\textbf{Fraud Detection}     & SHAP, Permutation            & Recursive Rule Extraction    & Activation Maps               & Sensitivity \& Perturbation  \\ \hline
\textbf{Risk Assessment}     & Gradient-Based               & Distillation Methods         & Concept-Based Probing         & Minimal Pair Analysis        \\ \hline
\textbf{Algorithmic Trading} & Feature Importance           & Knowledge Distillation       & Representation Engineering    & Market Stress Testing        \\ \hline
\end{tabular}%
}
\caption{Summary of the financial applications, mapped to interpretability techniques}
\label{tab:financial_use_cases}
\end{table}

\section{Mechanistic Interpretability - Concepts And Tools}
\label{sec:techniques}

Mechanistic interpretability seeks to reverse-engineer neural networks by identifying the precise computational pathways contributing to decision-making. Unlike other explainability methods, which analyze models after predictions are made, mechanistic interpretability aims to understand the internal mechanisms at a fine-grained level. This approach is crucial for ensuring AI safety, reliability, and transparency, particularly in high-stakes applications such as finance and healthcare.

A fundamental challenge in mechanistic interpretability is that modern AI models, particularly \emph{transformers}, store far more information than they have neurons or activations available. This compression of knowledge leads to a phenomenon called \textbf{superposition}, where multiple independent concepts are stored within the same activation space. As a result, individual neurons often become \textbf{polysemantic}, meaning they activate for multiple, seemingly unrelated features, making direct interpretation difficult.\footnote{A \emph{feature} in a neural network represents a distinct piece of learned information encoded in its internal activations.} These concepts were formalized in the influential research by \emph{Anthropic’s Transformer Circuits} team, particularly in their article on \emph{Superposition} \citep{elhage2022}.

Sparse autoencoders (SAEs) are among the most promising tools for representing disentangled and interpretable concepts. Given their importance and the focus of our experiments, we begin this section with an overview of SAEs, followed by several observation and intervention-based techniques in mechanistic interpretability, including methods such as Logit Lens and Attribution Patching.

We will explore these tools and concepts through practical experiments described in Section~\ref{sec:use_cases_without_sae} and Section~\ref{sec:sae_use_cases}. These include the use of Logit Lens and Attribution Patching on GPT-2 models, and sparse autoencoders on Gemma to reveal and manipulate interpretable features relevant to financial tasks.

\subsection{Sparse Autoencoders }
\label{sec:sae_basic}

Large Language Models (LLMs) are trained to maximize their efficiency, so they compress information into the fewest dimensions possible, allowing a limited number of neurons to encode a massive number of features. Instead of identifying a one-to-one mapping between neurons and concepts, researchers must use alternative techniques to uncover hidden structures within activations.

Sparse autoencoders (SAEs) are utilized to mitigate the issue of polysemantic neurons that activate for multiple, unrelated features. An SAE is a type of autoencoder that applies sparsity-promoting regularization to its hidden layer, encouraging each neuron to activate for a distinct feature. This method decomposes hidden layer activations into interpretable, sparse feature sets, ideally ensuring that each unit corresponds to a single monosemantic feature. This approach aligns with dictionary learning principles, where the objective is to represent data sparsely over a learned dictionary of features. The foundational work in this area, "Towards Monosemanticity", demonstrated the effectiveness of SAEs in disentangling features \citep{Bricken2023}.

Sparse autoencoders are two-layer multilayer perceptrons (MLP) with sparsity-promoting regularization applied to the hidden layer. The encoder in the SAE increases the dimensionality of the input data, \(x\), using a ReLU activation function, from the original dimension \(d_{\text{in}}\) to a higher dimension \(d_{\text{hid}}\). The encoder’s output is given by the equation:

\begin{equation}
\label{eqn:relu}
h = \text{ReLU}(W_{\text{enc}}x + b)
\end{equation}

where \(W_{\text{enc}}\) is the weight matrix and \(b\) is the bias term. The decoder reconstructs the input from original model hidden state \(x'\) using the equation:

\begin{equation}
\label{eqn:w_dec}
x' = W_{\text{dec}}h
\end{equation}

where \(W_{\text{dec}}\) forms the feature dictionary. Another way to describe \(W_{\text{dec}}\) is as decoder weights that define \emph{features} or \emph{concepts}. Each column of \(W_{\text{dec}}\) is a feature vector associated with a latent (i.e. hidden layer). This process encourages the model to represent the input data in terms of a sparse set of learned dictionary features, improving transparency and interpretability. The sparse autoencoder is trained to minimize a loss function, which includes a reconstruction term, \( \| x - x' \|_2^2 \), and a sparsity penalty term \( \| h \|_1 \), is the L1 norm, encouraging sparse activations in the hidden layer. The overall loss function is:

\[
L(x) = \| x - x' \|_2^2 + \alpha \| h \|_1
\]

where \( \alpha \) controls the strength of the sparsity penalty. 

\begin{figure}[h!]
\centering
\includegraphics[width=0.7\textwidth]{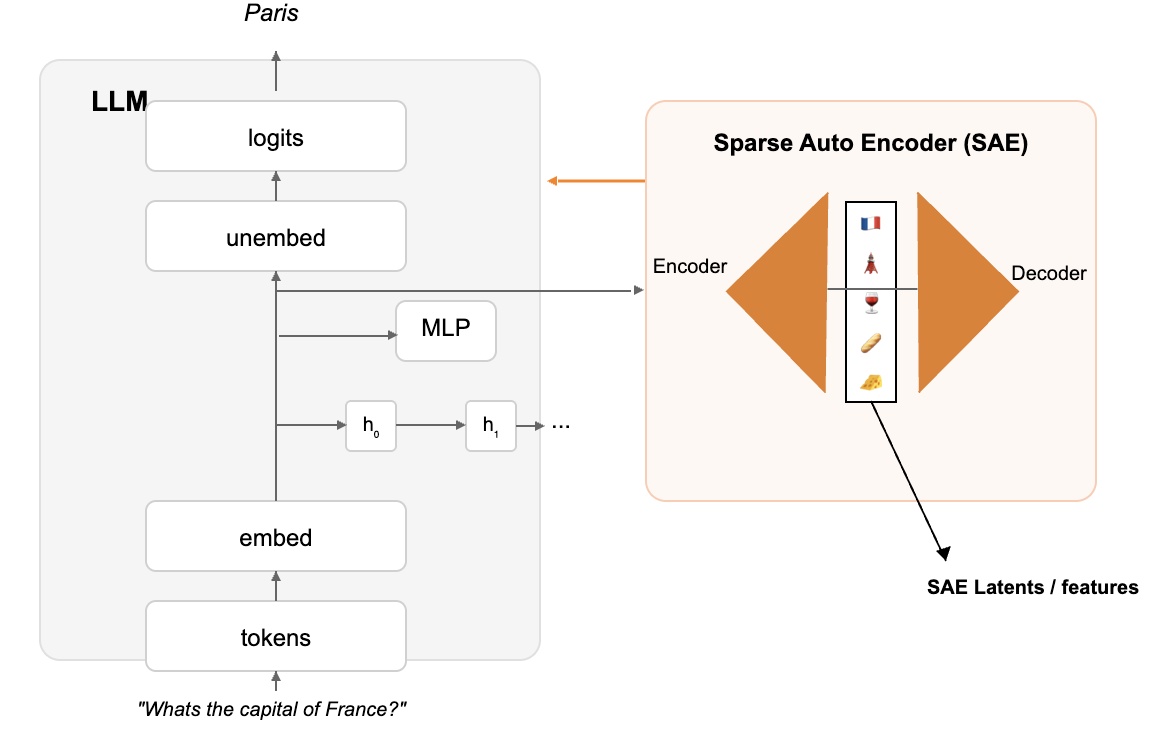}
\caption{Sparse Autoencoder}
\label{fig:sae_process}
\end{figure}

The diagram in Figure \ref{fig:sae_process} illustrates a process where an LLM (Large Language Model) takes tokens as input and converts them into embeddings, which are dense vector representations. These embeddings are processed through attention heads \(h_{0}, h_{1}...h_{n}\), which capture different relationships between words. The MLP (Multi-layer Perceptron) further refines these features before contributing to the final output logits. Additionally, a Sparse Autoencoder (SAE) is introduced to analyze activations from both the residual stream and MLP outputs, allowing it to disentangle complex feature representations into monosemantic units. For example, when asked about the capital of France, SAE features related to \textit{Flag of France}, \textit{Eiffel Tower}, \textit{wine} etc. activate, reflecting high-level conceptual associations. Sparse autoencoders learn to isolate such interpretable concepts by compressing activations into a small number of specialized latent features.

Section~\ref{sec:sae_use_cases} presents our research results based on sparse autoencoders, highlighting their applications in financial use cases such as trading, sentiment analysis, bias detection, and hallucination mitigation.

\subsubsection{Auto-Interpretation}  
\label{sec:auto_interp}
SAE features, though often interpretable, lack direct human-readable interpretations. To convert these features into understandable insights, in order to get the human-readable labels as per \href{https://www.neuronpedia.org/llama3.1-8b/steer}{Neuronpedia \cite{Neuronpedia2024}
} - a project initiated by Google's DeepMind, features are assigned labels like "references to London" by using an AI model to analyze the patterns in the texts where a feature is most strongly activated. These patterns are derived from the "Top Activations" of the feature, allowing the larger AI to identify and label the feature's focus based on the text it interacts with.

The usefulness increases as human-readable labels make it easier to understand, interpret, and explain the feature's role in the model, enhancing its interpretability and application.

Furthermore, the \textbf{magnitude of feature activation} in sparse autoencoders helps identify essential features, and enables performing numerical calculations, aiding in feature selection and statistical analysis for better decision-making.

Section~\ref{sec:sae_feature_discovery} shows our experiments related to the use of identifying the credit risk and bias using auto-interpretation of sparse features.

\subsubsection{Self-Interpretation} \label{selfinterp}
Another approach to get the human-readable interpretation is self-interpretability. Self-interpretability is a novel but intuitive concept described in \href{https://www.alignmentforum.org/posts/8ev6coxChSWcxCDy8/self-explaining-sae-features}{Self-explaining SAE features}  and originally conceived in \citep{chen2024selfieselfinterpretationlargelanguage}. The idea is to ask the model to describe a feature that is being steered via SAE autoencoder latents as defined $h$ in Equation \ref{eqn:relu}. 

Section~\ref{sec:sae_self_interpretability} demonstrates our use of self-interpretability to extract clusters of financial features from large language models.

\subsubsection{Steering Sparse Autoencoders}
\label{sec:steering_sparse_autoencoders}

In feature steering, SAEs' specific features or latents of interest are modified to artificially high or low values during the forward pass (see \href{https://transformer-circuits.pub/2024/scaling-monosemanticity/index.html\#appendix-methods-steering}{Methodological Details} for implementation details). Feature steering is remarkably effective at modifying model outputs in specific, interpretable ways. It can be used to modify the model’s demeanor, preferences, stated goals, and biases; to induce it to make specific errors.

Using the definition of SAE decoder reconstruction in equation \ref{eqn:w_dec}, we can modify sparse hidden state $h$ with a steering vector $\Delta{h}$.

\[
\hat{x}' = W_{\text{dec}}(h + \Delta{h})
\]

Sparse autoencoders (SAEs) described earlier have emerged as a powerful tool for interpretability. Instead of modifying raw activations, SAEs learn \textbf{disentangled representations} that allow targeted interventions in model behavior \citep{Bricken2023}.

Section~\ref{sec:feature steering} presents our research experiments demonstrating how feature steering enables modification of LLM's output related to credit risk and enhances performance in sentiment analysis.

\subsection{Mechanistic Interpretability Methods: Observation vs. Intervention}
\label{sec:mech_interp_types}

\begin{figure}[h]
  \centering
  \includegraphics[width=.7\linewidth]{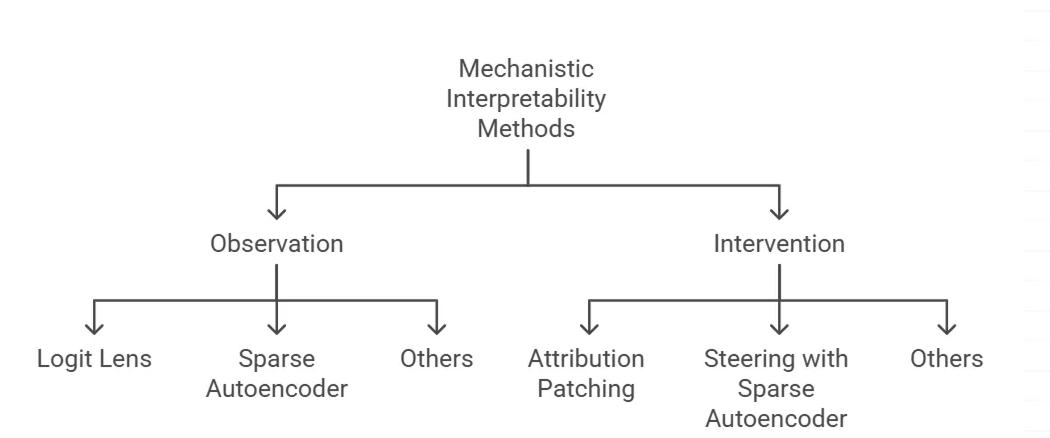}
  \caption{Interpretability Methods}
  \label{fig:Mechanistic Interpretability Methods}
\end{figure}

Mechanistic interpretability research is divided into two major categories as shown in Figure \ref{fig:Mechanistic Interpretability Methods}:
\begin{itemize}
    \item \textbf{Observation-Based Methods} – Analyze model activations without modifying the model.
    \item \textbf{Intervention-Based Methods} – Actively modify neural computations to establish causal relationships.
\end{itemize}

\subsection{Observation-based Methods}

Observation-based methods in mechanistic interpretability analyze a large language model's (LLM) internal activations without modifying their parameters. Sparse autoencoder-based feature detection helps reveal hidden patterns, showing how models organize and process information across layers. These techniques identify emerging features and the mechanisms driving predictions, enhancing interpretability.

They can be broadly categorized into different frameworks: \textbf{Logit Lens}, \textbf{Sparse Autoencoders}, and Other Methods such as structured probing and feature visualization. Given that Sparse Autoencoders were introduced in Section~\ref{sec:sae_basic}, this section covers Logit Lens and other observation-based techniques in mechanistic interpretability.

\subsubsection{Logit Lens}
\label{sec:logit_lens}
One of the fundamental approaches to observing neural activations is the \textbf{Logit Lens} which applies the next-word prediction at every layer. This enables researchers to analyze how confident a model is at predicting outputs at different layers, offering a glimpse into the step-by-step evolution of predictions. Mathematically, for an intermediate activation $h_\ell$ at layer $\ell$, the Logit Lens maps it to vocabulary space using the final model output projection matrix $W_{\text{out}}$ and bias $b_{\text{out}}$ from the pre-trained model:

\[
\hat{y}_\ell = W_{\text{out}} h_\ell + b_{\text{out}}
\]

where $\hat{y}_\ell$ represents the logits at layer $\ell$. Here, $W_{\text{out}}$ and $b_{\text{out}}$ are directly inherited from the model's original language modeling head, trained to predict the next token at the final layer. For example, in GPT models, these parameters are optimized to map the last hidden state to vocabulary logits during training. If earlier layer logits closely resemble final outputs, it suggests that the model forms predictions incrementally rather than exclusively at the last layer.

Section~\ref{subsec:logit_lens} demonstrates our use of Logit Lens to analyze token prediction dynamics across layers in financial sentence prompts.

\subsubsection{Other Observational Techniques}

\textbf{Probing for Features:} Feature probing involves training external classifiers on intermediate activations to test whether a model encodes specific properties. Simple linear probes \citep{alain2016} examine linear separability, while structured probes attempt to identify high-level concepts.

\textbf{Feature Visualization:} Feature visualization techniques generate synthetic inputs that maximally activate neurons, revealing their sensitivity to different patterns. Activation Maximization \citep{zeiler2014} iteratively adjusts inputs to highlight neuron-preferred features.

\textbf{Attention Analysis:} Observing attention distributions can offer interpretability insights in transformer models. Attention Lens \citep{sakarvadia2023} maps attention head activations to vocabulary tokens to uncover hidden feature relationships.

\subsection{Intervention-Based Methods}

Interventional techniques provide a causal framework for understanding and modifying model behavior by perturbing activations or weights and measuring the resulting effects. 

Intervention techniques in mechanistic interpretability include methods such as activation patching, steering sparse autoencoders, and causal abstraction. Given that steering sparse autoencoders was already covered in Section~\ref{sec:steering_sparse_autoencoders}, we now turn to the remaining intervention techniques in the following section.

\subsubsection{Attribution Patching}
\label{sec:attribution_patching}
One of the most widely used interventional methods is attribution patching, which involves modifying input data to determine how specific features contribute to the model’s predictions. This technique helps to attribute the model’s decisions to particular components of the input, providing insights into which aspects of the data are most responsible for the model's outputs. Attribution patching can be linked to techniques such as feature attribution \citep{Sundararajan2017}, and Shapley value-based analysis \citep{Lundberg2017}.

Attribution patching involves the following key steps:

\begin{enumerate}
    \item Running the model with a \textbf{clean input} to obtain the baseline predictions.
    \item Modifying or \textbf{corrupting} certain parts of the input, effectively "patching" or removing specific features.
    \item Running the model again on the modified input and observing the changes in the output.
    \item The \textbf{attribution of importance} is determined by measuring how much the modified input alters the predictions, thereby indicating the significance of each part of the input.
\end{enumerate}

\begin{figure}[!h]
\centering
\includegraphics[width=0.6\textwidth]{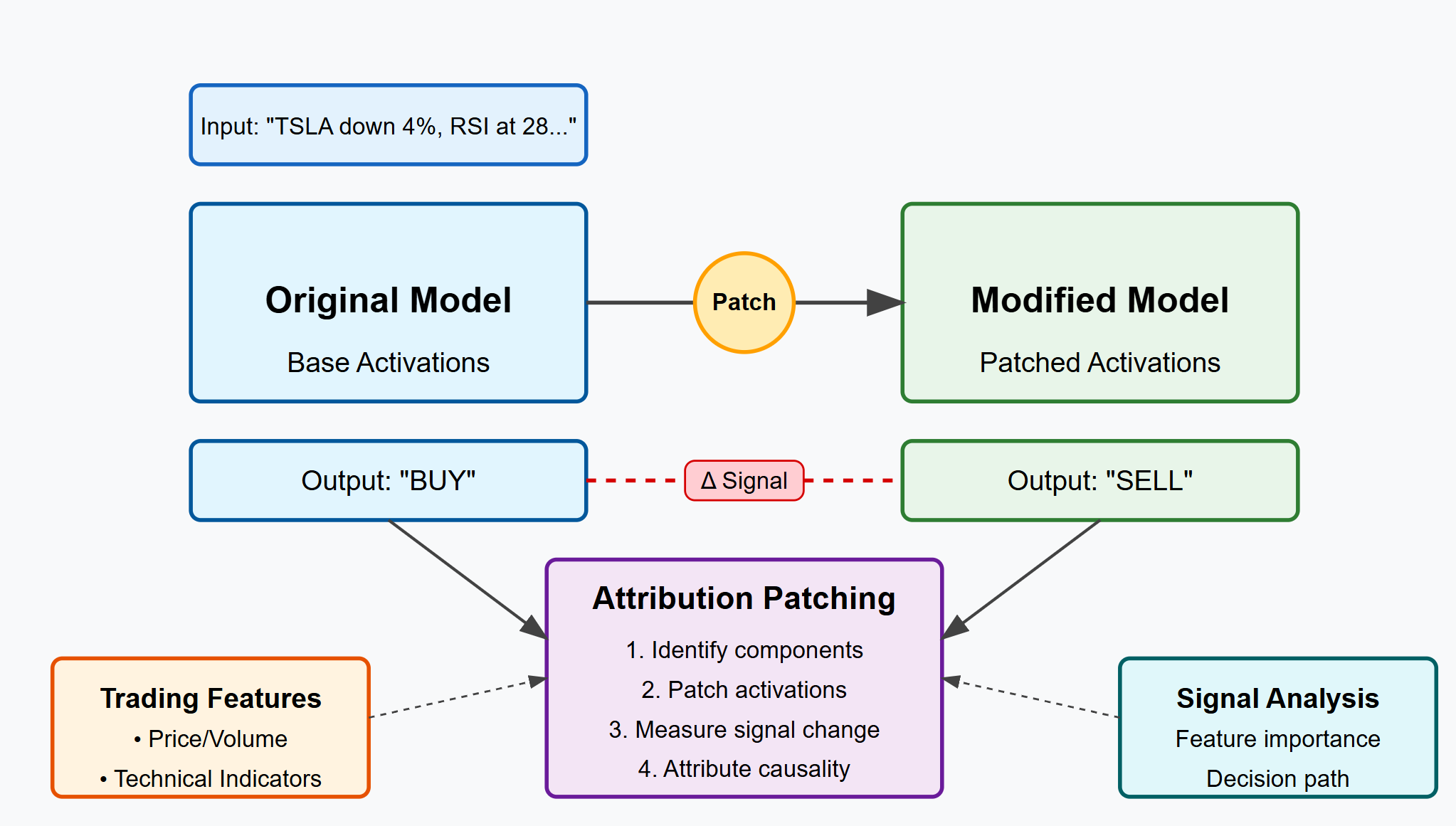}
\caption{Attribution patching analysis}
\label{fig:attribution_patching}
\end{figure}

As illustrated in Figure \ref{fig:attribution_patching}, this process compares an original model (left) correctly outputting \textbf{buy}, when evaluating Tesla stock data against a modified version (right) with patched activations, producing \textbf{sell} with the central intervention point represented by the \textbf{Patch} node that modifies specific neural components. The methodology follows a structured four-step attribution patching approach: identifying relevant components (price/volume features, technical indicators), implementing targeted activation patches, measuring resulting signal differences, and establishing causal relationships between internal mechanisms and trading decisions. This rigorous approach extends beyond correlational analysis to determine which components are causally necessary for specific computational pathways, enabling researchers to trace information flow through trading algorithms and precisely identify the mechanisms responsible for particular buy or sell recommendations.

Section~\ref{subsec:attr_patch_experiment} demonstrates our experiment on how attribution patching reveals influential heads and layers in financial sentence predictions using clean and corrupted prompts.

\subsubsection{Other Intervention Techniques}

In addition to attribution patching and SAE steering, several alternative intervention techniques exist:

\begin{itemize}
    \item \textbf{Causal Abstraction}: Maps neural computations onto human-understandable structures, making high-level AI reasoning transparent \citep{xia2024neural}.
    \item \textbf{Resample Ablation}: Replaces activations with \textbf{r}andomized values to test the stability of circuits in transformers \citep{Zhang2023}.
    \item \textbf{Feature Injection}: Introduces synthetic activations to study how specific concepts influence outputs \citep{lin2024}.
\end{itemize}

\section{Applications in Financial Services}
As mentioned in Section~\ref{sec:mech_interp_types}, Mechanistic Interpretability techniques can be broadly classified into \textbf{observation-based methods} and \textbf{intervention-based methods}. Observation techniques passively analyze how a model processes information, whereas intervention techniques actively modify its activations to test causal effects.

Although these methods, such as sparse autoencoder-based feature analysis, activation steering, and attribution techniques, show promise in enhancing transparency, bias mitigation, and compliance, they are not yet widely known or implemented in real-world financial systems. They represent potential applications that could improve AI-driven decision-making in areas like credit scoring, fraud detection, and algorithmic trading.

Some potential use cases for applying mechanistic interpretability techniques in finance are mentioned below and their detailed applications are illustrated in Table \ref{tab:InterpFinance}.

\subsection{Observation-Based}

\subsubsection{Feature Discovery And Identification}

Observation-based techniques like \textbf{Logit Lens} mentioned in Section~\ref{sec:logit_lens} and \textbf{sparse autoencoders} mentioned in Section~\ref{sec:sae_basic} are great tools to identify the features of the model. The Logit Lens applies the next-word prediction every layer, showing how the model's guesses evolve and become more refined across layers. This helps understand model decision-making in areas such as credit scoring and fraud detection. Sparse autoencoders, on the other hand could recover monosemantic features from a small one-layer transformer. Together, these methods increase model transparency and risk mitigation. 

\subsubsection{Feature Activation-Based}

The magnitude of feature activation of SAE essentially provides a quantifiable measure of the model's focus on specific aspects of the input or output, enabling a deeper understanding of its decision-making process and facilitating diverse applications across various domains. By analyzing activation patterns, models can identify inconsistencies in generated content, detect biases in decision-making, and extract insights for trading strategies. In hallucination detection, feature activation helps flag potentially incorrect outputs. For bias detection, it highlights sensitive attributes that may influence model decisions, ensuring fairer outcomes. In trading strategies, feature activation aids in mapping financial news sentiment to market movements, enabling more informed predictions. This approach enhances transparency and reliability in AI-driven financial analysis.

\subsection{Intervention-Based}

\subsubsection{Attribution Patching}

As mentioned in Section~\ref{sec:attribution_patching}, attribution patching tests the impact of neuron activations by modifying specific layers in a model. A typical example is using corrupted and clean financial reports—where a model predicts bankruptcy for one company based on a report, and then a corrupted version of the same report is used to analyze which activations determine the final prediction. This method helps in identifying which components of a model contribute most significantly to financial distress predictions and fraud detection.

\subsubsection{Steering With SAEs}
Feature steering as described in Section~\ref{sec:steering_sparse_autoencoders} modifies model behavior internally, reducing reliance on prompt engineering and fine-tuning. In sentiment analysis, for example, it can direct models toward specific risk assessments or investment strategies. This enables precise AI-driven financial decisions without costly retraining. However, fine-tuning is necessary when introducing new knowledge since feature steering only adjusts existing activations rather than incorporating external information.

\begin{table}[h!]
\centering
\resizebox{\textwidth}{!}{%
\begin{tabular}{|l|p{2.5cm}|p{3.5cm}|p{6.5cm}|}
\hline
\textbf{Category} & \textbf{Techniques} & \textbf{Potential Finance Applications} & \textbf{Application Details} \\ \hline
\textbf{Feature Discovery and Identification} & Logit Lens, Sparse Autoencoders & Credit Scoring, Fraud Detection, Regulatory Compliance & Track prediction evolution across layers using Logit Lens to understand model decisions; apply sparse autoencoders (SAEs) to extract monosemantic financial features, increasing model transparency and risk mitigation. \\ \hline
\textbf{Feature Activation-Based} & Sparse Autoencoder-Based Analysis & Sentiment Analysis, Trading Strategy, Hallucination Detection, Bias Detection & Analyze feature activation patterns to detect inconsistencies, biases, and map financial news sentiment to market movements, aiding in informed predictions and ensuring accurate, fair model decisions. \\ \hline
\textbf{Attribution Patching} & Activation Patching & Bankruptcy Prediction, Fraud Detection, Financial Reporting & Test the impact of neuron activations using corrupted and clean financial reports to analyze which activations contribute most to financial distress predictions and fraud detection. \\ \hline
\textbf{Feature Steering with SAEs} & Activation Steering & Trading Strategy Optimization, Regulatory Explanation & Modify model outputs in a specific, interpretable way by steering SAEs to high or low values; steer models toward specific risk assessments or investment strategies for more accurate financial decisions without retraining. \\ \hline
\textbf{Feature Steering as a Fine-Tuning Alternative} & Activation-Based Feature Refinement & Fraud Detection, Portfolio Risk Analysis & Replace expensive model fine-tuning by adjusting internal activations for real-time adaptation of fraud detection models and portfolio stress-testing. \\ \hline
\end{tabular}%
}
\caption{Mechanistic Interpretability Techniques in Finance}
\label{tab:InterpFinance}
\end{table}

\section{Experimental Results (Without Sparse Autoencoders)}
\label{sec:use_cases_without_sae}

The experiments presented in the following sections build directly on the foundational concepts discussed earlier, applying them in practical financial scenarios. For Logit Lens and Attribution Patching, we primarily use GPT-2 models to analyze token probability evolution and activation importance across layers. For Sparse Autoencoder-based experiments, Google’s Gemma models are primarily employed to uncover and steer interpretable features. Together, these experiments demonstrate how different interpretability techniques can be systematically applied to understand and enhance model behavior in financial applications.

\subsection{Logit Lens}
\label{subsec:logit_lens}

In this experiment, \textbf{Logit Lens} is employed to track how probabilities shift across layers, allowing for a clearer understanding of which parts of the model influence specific financial predictions. We used the \textbf{GPT-2} model for this analysis.

\begin{figure}[h]
  \centering
  \includegraphics[width=.7\linewidth]{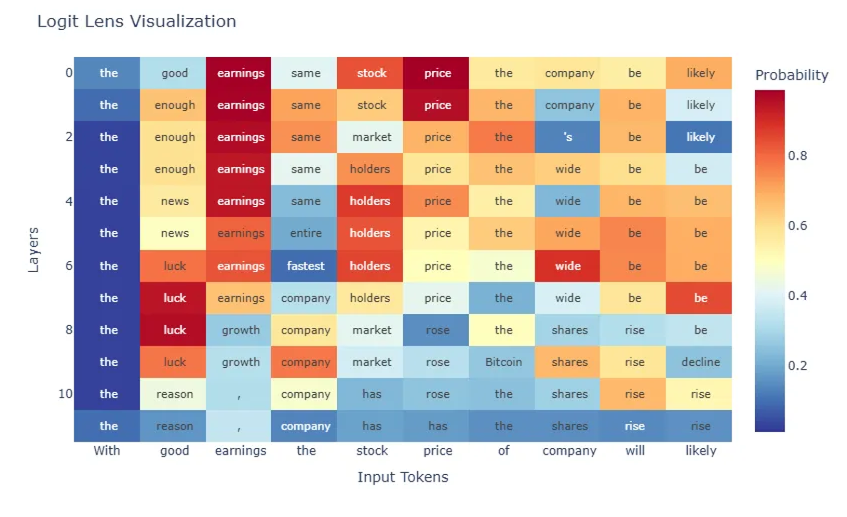}
  \caption{Logit Lens heatmap showing predicted token probabilities at each layer for a financial prompt}
  \label{fig:Logit}
\end{figure}

First, we used the Logit Lens to generate a heatmap as shown in Figure \ref{fig:Logit} illustrating the model's predicted token probabilities at each layer for a given financial prompt. The specific sentence used was:

\begin{quote} \textit{"With good earnings the stock price of company will likely"} \end{quote}

In the heatmap, the X-axis represents each input token from the prompt, and the Y-axis represents the model layers from 0 to 10. Each cell shows the model's most probable next word prediction given the preceding context at that particular layer, with color intensity indicating the associated probability. The probability reflects the model's confidence that the word it predicts is the correct next token given the previous tokens. This visualization allows us to observe how the model’s guesses evolve and become more refined as the input flows deeper into the network.

\begin{figure}[h]
  \centering
  \includegraphics[width=.7\linewidth]{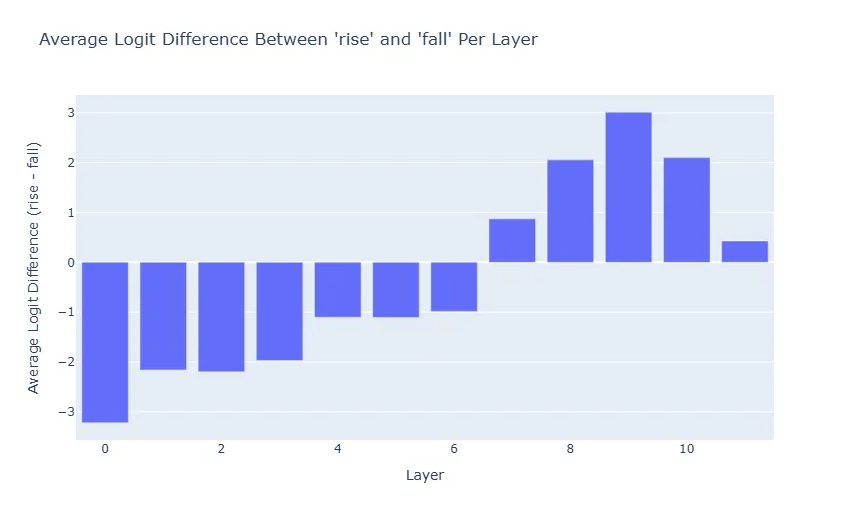}
  \caption{Logit difference chart between the words "rise" and "fall" across different layers}
  \label{fig:Logit_Diff}
\end{figure}

Figure \ref{fig:Logit_Diff} shows the average logit difference between the words \textbf{rise} and \textbf{fall} across different layers of the model. The negative logit difference in the earlier layers (0-4) indicates that the model initially prefers \textbf{fall} over \textbf{rise} when processing the input. However, as the model progresses through layers 5 to 10, there is a shift toward a higher preference for \textbf{rise}, particularly in the deeper layers (9-10), where the difference becomes more pronounced. This suggests that the deeper layers are refining the model's judgment, leading to a stronger final decision in favor of \textbf{rise}, showing the model's increased confidence in this choice as it processes the input.

In addition to this, we also analyzed the reverse scenario, where the model's reasoning is tested under more negative financial conditions. The second prompt used was:

\begin{quote} \textit{"With recession risk the stock price of company will likely"} \end{quote}

\begin{figure}[h]
  \centering
  \includegraphics[width=.7\linewidth]{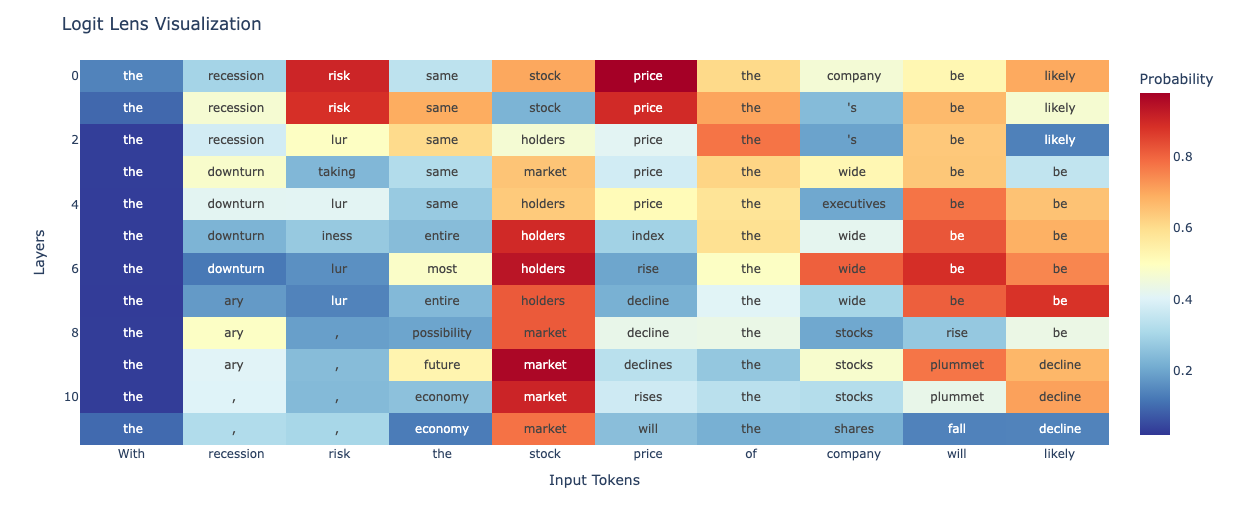}
  \caption{Logit Lens heatmap showing predicted token probabilities at each layer for a financial prompt}
  \label{fig:Log_Reverse}
\end{figure}

The corresponding heatmap is shown in Figure \ref{fig:Log_Reverse}. As with the previous figure, the X-axis represents the input tokens and the Y-axis represents the model layers. In this case, the model's predictions shift differently, reflecting the more negative context embedded in the prompt. We observe that the model increasingly favors negative outcomes such as "fall" and "decline" as it moves through the layers, indicating that the model adjusts its expectations appropriately in response to recession-related terms. Again, the color intensity reflects the probability assigned to each predicted next word, allowing for visualization of how confident the model becomes at each processing stage. 

Logit attribution can enhance model trustworthiness by providing clear explanations for AI-driven decisions, reducing bias, and ensuring fairness in automated decision-making systems.

\subsection{Attribution Patching}
\label{subsec:attr_patch_experiment}

As mentioned in Section~\ref{sec:attribution_patching}, attribution patching involves running the model on both clean and corrupted prompts and systematically replacing specific activations from the corrupted run with activations from the clean run. If replacing an activation significantly increases the likelihood of the correct response (e.g., recognizing the correct stock price movement based on revenue growth), it indicates that the patched activation is important for task completion.

In this study, the experiment involves using a series of prompts that test the model’s ability to predict outcomes based on varying financial scenarios. 

Each prompt is evaluated separately through its clean and corrupted version, and the patching results are aggregated across prompts to generate overall insights.

The clean prompt follows a structure where a business or financial scenario explicitly tests the model’s understanding of key principles such as revenue growth, financial risk, and investor sentiment:

\begin{quote}
    \textit{If a company's revenue grows steadily, its stock price will likely increase.}
\end{quote}

The corrupted prompt maintains the structure but modifies a key variable to challenge the model’s understanding:

\begin{quote}
    \textit{If a company's revenue grows steadily, its stock price will likely decrease.}
\end{quote}

Here, the expected relationship between revenue growth and stock price increase is disrupted, making it more difficult for the model to apply its usual reasoning.

\begin{table}[h!]
    \centering
    \scriptsize 
    \resizebox{\textwidth}{!}{ % Rescales table to fit within text width
        \begin{tabular}{|p{7cm}|p{3cm}|p{3cm}|}
            \hline
            \textbf{Prompt} & \textbf{Correct} & \textbf{Incorrect} \\ \hline
            If a company's revenue grows steadily, its stock price will likely & 'increase' & 'decrease' \\ \hline
            A company with high debt and low cash flow is considered financially & 'risky' & 'stable' \\ \hline
            During periods of economic recession, consumer spending typically & 'decrease' & 'increase' \\ \hline
            When interest rates rise, bond prices generally tend to & 'fall' & 'rise' \\ \hline
            A company with a diversified portfolio of investments is less & 'risky' & 'safe' \\ \hline
            If a company consistently reports strong earnings, its investors are likely to be & 'satisfied' & 'dissatisfied' \\ \hline
            When inflation increases, the purchasing power of money & 'decreases' & 'increases' \\ \hline
            A company with a strong brand reputation often commands a higher & 'price' & 'discount' \\ \hline
        \end{tabular}
    }
    \caption{Examples of financial prompts used in the attribution patching experiment}
\end{table}

\begin{figure}[h!]
\centering
\includegraphics[width=0.7\textwidth]{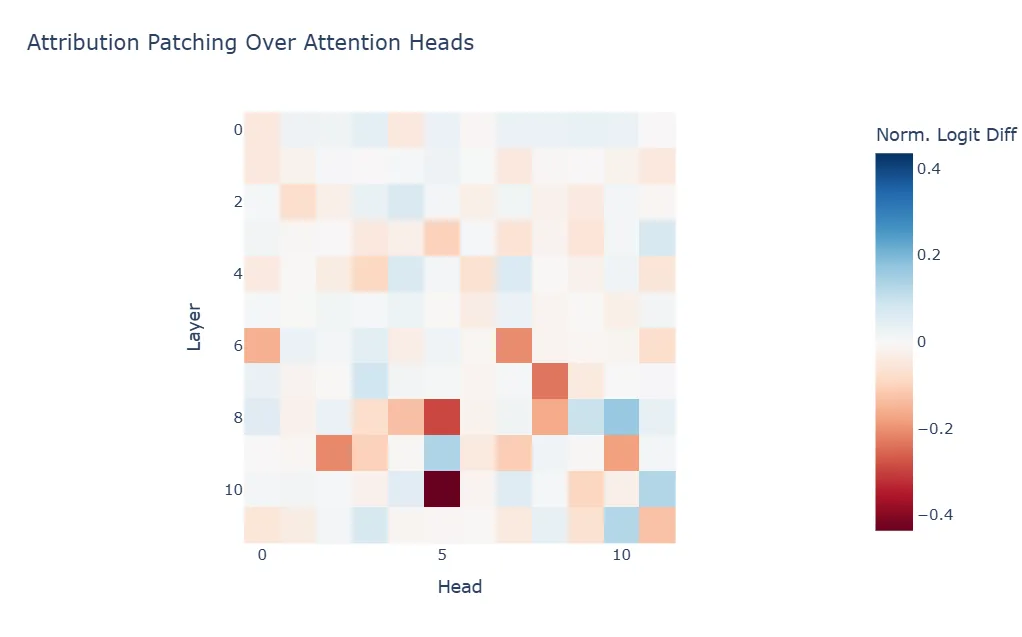}
\caption{Heatmap visualizing attribution patching over attention heads in the model}
\label{fig:heatmap_attr}
\end{figure}

The heatmap in Figure \ref{fig:heatmap_attr} visualizes attribution patching over attention heads, indicating the impact of specific layers and heads on model predictions. The X-axis represents attention heads and the Y-axis represents layers within the GPT-2 Small model. GPT-2 Small has 12 layers and 12 attention heads per layer, with each head attending to different parts of the input context during prediction. The normalized logit difference scale (ranging from negative to positive values) reflects which activations reinforce or suppress certain outputs.

\href{}{}
\subsection*{Key Observations:}

\begin{itemize}
    \item \textbf{Layer-Wise Influence:} The most critical activations appear in layers 8 and 10, where specific heads strongly influence the final model decision. Layer 6 shows moderate suppression, suggesting interference from unrelated features rather than contributing positively to the decision-making process.
    \item \textbf{Attention Head Contributions:} Attention heads in layers 8 and 10 (e.g., 8.5, 10.5, 10.7) significantly reinforce correct predictions, while heads in layer 9 (e.g., 9.4, 9.6) exhibit negative contributions, potentially introducing noise or conflicting signals.
    \item \textbf{Feature Steering Potential:} The presence of strong negative contributions (dark red regions) in certain heads suggests that activation steering or selective suppression could improve interpretability and decision reliability in financial applications.
\end{itemize}

This study highlights the dominance of layers 8 and 10 in shaping final predictions, with certain attention heads reinforcing or suppressing outputs.  Financial-theme related circuits identified through this method might tend to activate similar heads across related prompts, suggesting partial consistency in the model’s internal reasoning for financial tasks. This approach can trace model decision-making and pinpoint interpretable patterns that can guide targeted model improvements or risk monitoring.

%----------------------------------------------------
\section{Experimental Results Using Sparse Autoencoders}
\label{sec:sae_use_cases}
In this section, we explore Sparse Autoencoder-based applications, taking a deeper step into feature-level interpretability. We first discuss how to label features through Feature Discovery and Identification, followed by experiments on Prediction, Bias Reduction, and Hallucination Reduction. Most of the experiments are based on Sparse Autoencoders trained on Google's Gemma models, and the techniques demonstrated can be broadly extrapolated across different financial applications.

\subsection{Feature Discovery and Identification}
\label{sec:sae_feature_discovery}
Using sparse autoencoders for feature discovery can help identify and activate the most relevant features, improving interpretability and efficiency. By mapping neural behavior, sparse autoencoders can aid in targeted fine-tuning and support prompt engineering, structured knowledge representation, and improved decision-making.

\subsubsection{Self-Interpretability}
\label{sec:sae_self_interpretability}
As briefly described in \ref{selfinterp}, we are able to interpret and explain SAE latent representation of features via SAE steering. We simply prompt a model to describe a generic concept,  steer particular tokens via SAE latent and see a steered concept being described. An algorithm to generate a list describing all the features is the following: We steer (increase activation) one feature at a time across all the features available in our SAE, while providing the LLM with the following prompt: 

    \textit{What is the meaning of the word "X"?The meaning of the word "X" is}

The steering effect is focused on the word X in the prescribed prompt. The result is a complete list of SAE induced features described in plain English. Our experiments are using Gemma Scope
SAEs \citep{lieberum2024gemmascopeopensparse} trained on Gemma-2B \citep{gemmateam2024gemma2improvingopen} with dictionary size 16,384. A small sample of discovered finance related features is in Table \ref{tab:feature_description}.

\begin{table}[h]
    \centering
    \renewcommand{\arraystretch}{1.2} % Adjust row height for better readability
    \begin{tabular}{|c|p{12cm}|}
        \hline
        \textbf{Feature Id} & \textbf{LLM-generated feature description} \\
        \hline
        471 & The seriousness, credit, or financial risk associated with an entity. It is used by investors, lenders, and other users of financial information to assess the risk and make informed decisions. A formal assessment of the creditworthiness or financial stability of an organization or individual. \\
        \hline
        2691 & To give, yield, or produce. It can also refer to the result of something or a thing that is produced. It is often used in the sense of introducing, facilitating, or promoting something, typically in relation to providing information, knowledge, or entertainment. \\
        \hline
    \end{tabular}
    \caption{LLM-generated feature descriptions}
    \label{tab:feature_description}
\end{table}

\begin{figure}[htbp]
    \centering
    \includegraphics[width=0.55\textwidth]{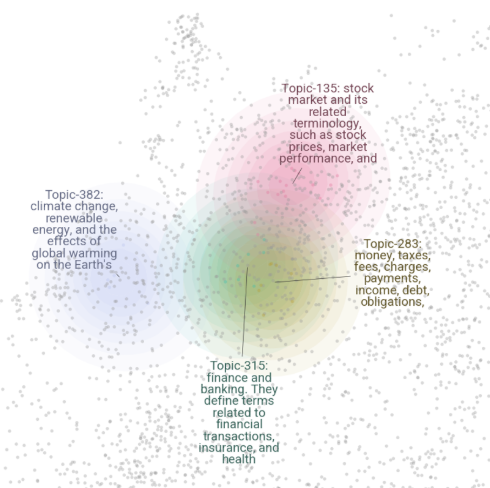}
    \caption{Feature clustering visualization highlighting finance-related topics}
    \label{fig:feature_clusters}
\end{figure}

 We are also able to create semantic clustering of all LLM self-interpretable generated features. LLMs are known to have redundancies built-in, so naturally some features are related to each other. Using decoder latents $W_{\text{dec}}$ from our SAE as embeddings provided to a clustering algorithm, we are able to see a hierarchy of emerging features. We follow the clustering algorithm described in \citep{grootendorst2022bertopicneuraltopicmodeling} where each SAE feature vector is measured via a vector similarity distance function that eventually forms a cluster. Figure~\ref{fig:feature_clusters} presents a visualization of clustered features based on their semantic similarity. The highlighted clusters indicate key financial topics, showing how related terms group together. The finance-related topics identified include:

\vspace{-0.5em}
\begin{itemize}
    \item \textbf{Topic-135:} Stock market and related terminology, including stock prices and market performance.
    \item \textbf{Topic-283:} Money, taxes, fees, charges, payments, income, debt, and obligations.
    \item \textbf{Topic-315:} Finance and banking, covering financial transactions, insurance, and health.
\end{itemize}
\vspace{-0.5em}

These clusters align with features identified in Table~\ref{tab:feature_description}, particularly Feature 471, which focuses on credit risk, and Feature 2691, which relates to financial outputs. The clustering helps interpret how financial concepts are grouped and analyzed within the dataset.

\subsubsection{Example 1: Credit Risk Related Features}

\begin{figure}[H]
    \centering
    \includegraphics[width=0.4\textwidth]{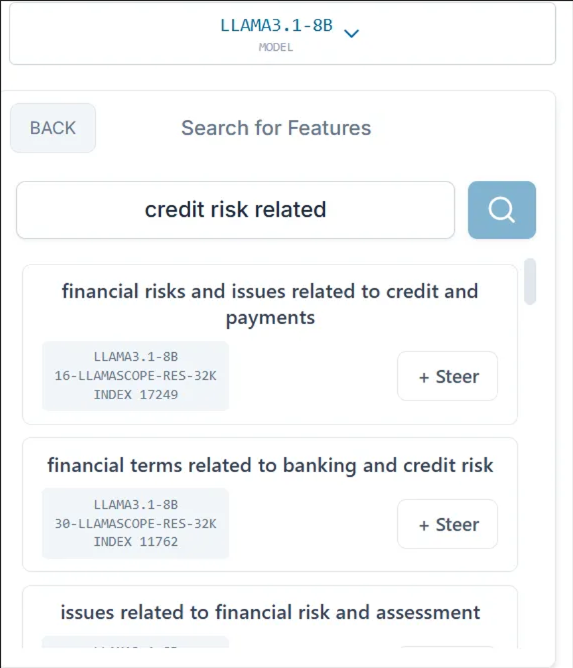}
    \caption{Neuronpedia search results for \textit{credit risk related features}}
    \label{fig:credit_risk_features}
\end{figure}

Neuronpedia is a DeepMind-supported platform that helps users explore and steer AI models by identifying specific capabilities based on search queries. As shown in Figure \ref{fig:credit_risk_features}, the phrase \textbf{credit risk related} was searched, and the identified features include \textit{financial risks and issues related to credit and payments}, \textit{financial terms related to banking and credit risk}, and \textit{issues related to financial risk and assessment}. These features reflect various aspects of credit risk, such as assessing financial risks in payment systems, understanding banking and credit-related terminology, and evaluating overall financial risk.

\subsubsection{Example 2: Bias Related Features}

\begin{figure}[h]
    \centering
    \includegraphics[width=0.5\textwidth]{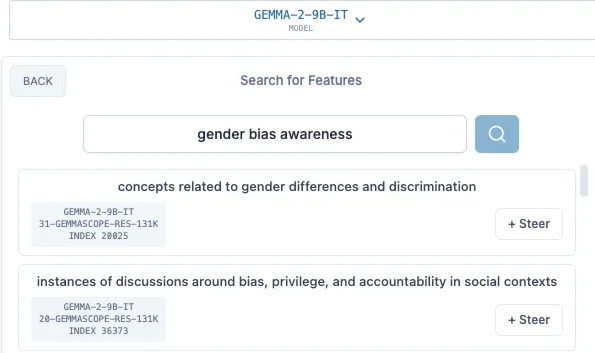}
    \caption{Neuronpedia search results for \textit{gender bias awareness features}}
    \label{fig:gender_bias_features}
\end{figure}

As shown in Figure \ref{fig:gender_bias_features}, the keyword \textbf{gender bias awareness} was searched using the Gemma-9B model, a Sparse Autoencoder corresponding to the Gemma model. This approach identified features related to \textit{concepts concerning gender differences and discrimination} and \textit{instances of discussions on bias, privilege, and accountability in social contexts}. These features reflect various aspects of gender bias, such as recognizing discrimination in different settings and discussions on social accountability regarding bias and privilege.

The Gemma-9B model is part of the Gemma Scope project, which provides a suite of JumpReLU Sparse Autoencoders trained on various layers of Gemma 2 models, including the 9B parameter variant. This project enhances the interpretability of large language models by decomposing their activations into interpretable features.

\subsection{Feature Activation-Based Experimental Results}

As mentioned earlier, the magnitude of feature activation provides a quantifiable measure of the model’s focus on specific input or output aspects, enabling deeper understanding of decision-making. In this section, we use feature activation magnitude for prediction, bias detection, and identifying hallucinations.

\subsubsection{Example: Prediction}
In this example, we explore how sparse autoencoder (SAE) feature activations can be leveraged to develop interpretable prediction models for short-term stock price movements (up or down) based on financial news headlines. Using activations from the SAE of the Gemma-2B model and a labeled dataset of financial news headlines with corresponding price movements, we performed feature discovery to identify the most relevant features from the dataset. Once the key features were identified, we trained a decision tree classifier to find the optimal combination of features and evaluated its performance using the F1 score.

The decision tree classifier, trained on this data, is visualized to show how different features contribute to predicting stock price movements (up or down). The tree reveals that certain features, such as financial terms and stock ticker symbols, play a more significant role in the prediction, while others have less influence. The intuition behind the tree is that it splits the data based on these features, progressively narrowing down predictions about whether a stock’s price will increase or decrease. 

\begin{figure}[H]
    \centering
    \includegraphics[width=0.8\textwidth]{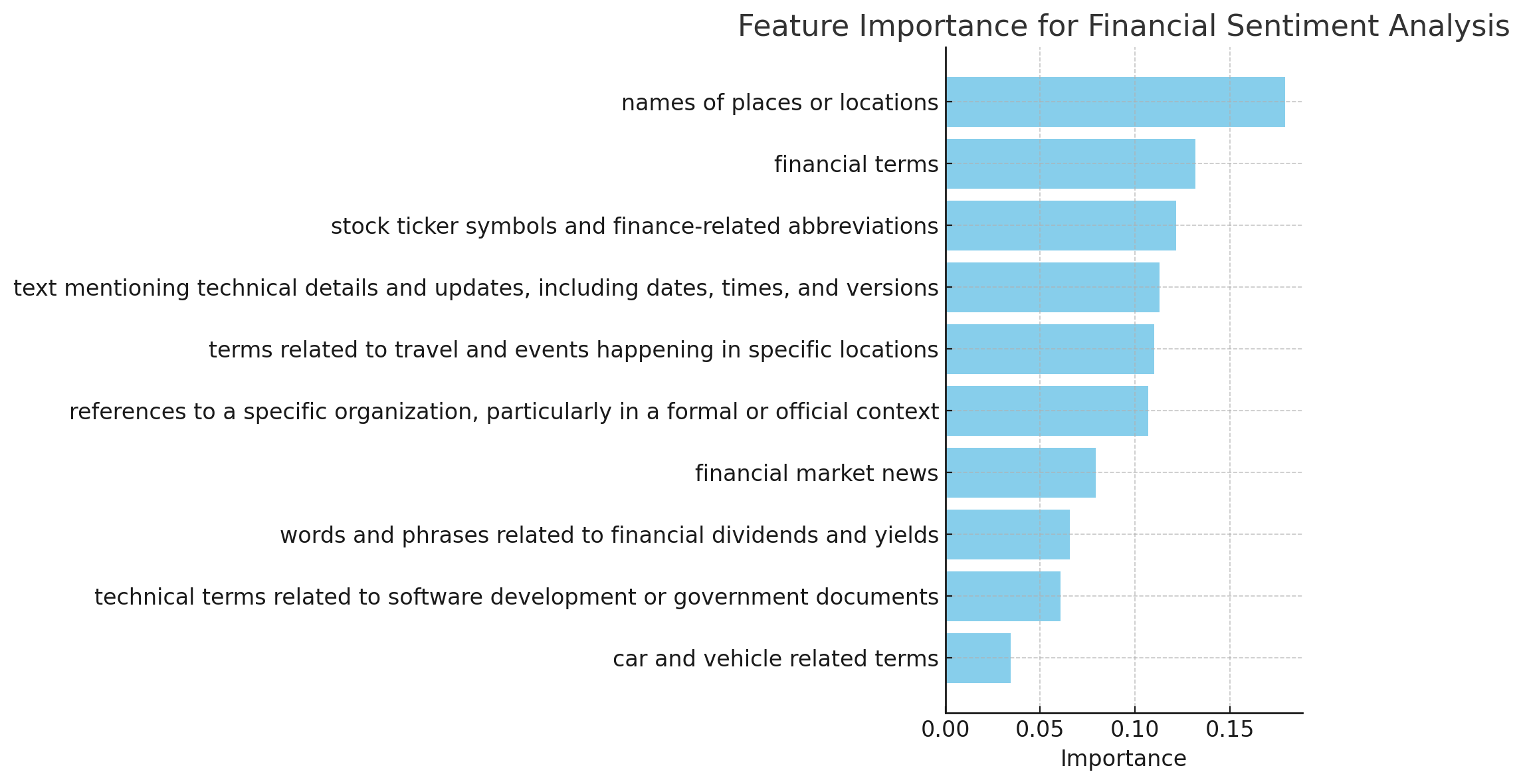}
    \caption{Feature Importance For Price movement}
    \label{fig:feature_importance}
\end{figure}

Figure \ref{fig:feature_importance} shows the feature importance chart, highlighting the relevance of features such as names of places or locations, financial terms, and stock ticker symbols. These features are intuitive and capture key factors such as financial terms, market trends, company-specific news etc.

\begin{figure}[H]
    \centering
    \includegraphics[width=1\textwidth]{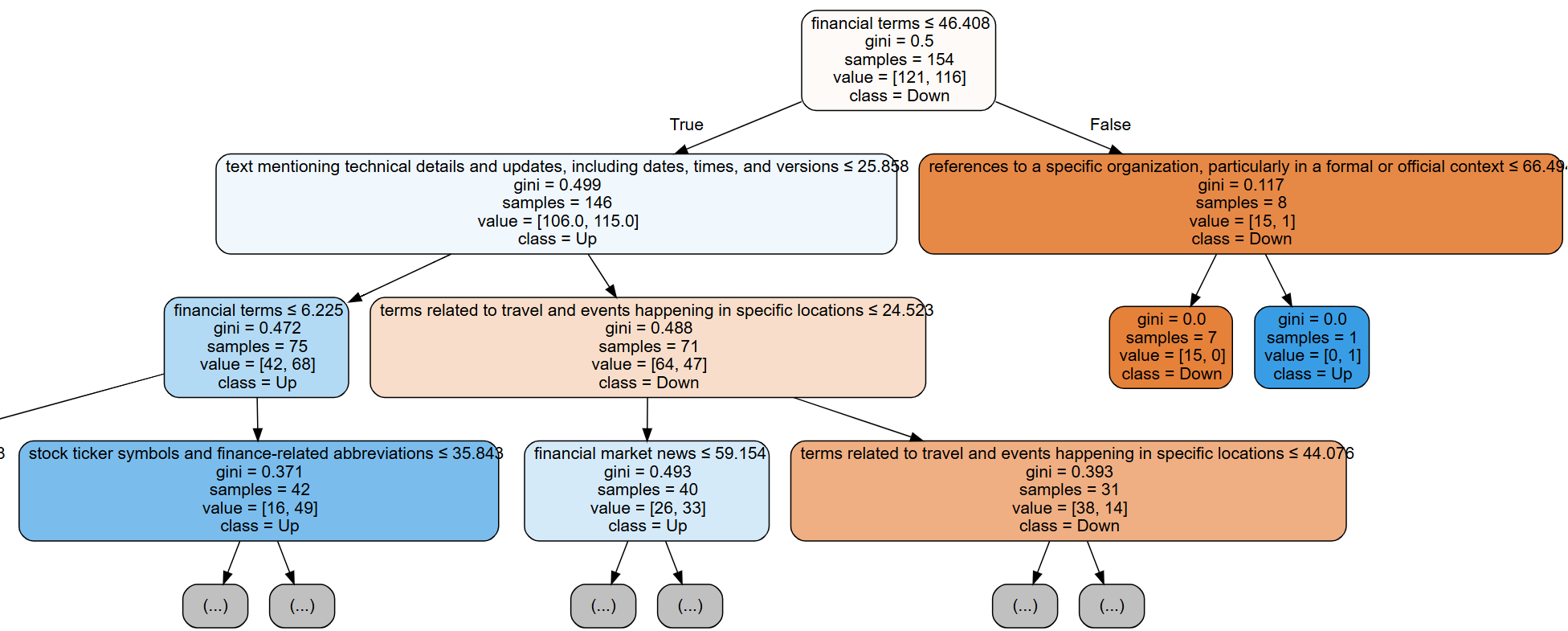}
    \caption{Decision tree visualization for price movement}
    \label{fig:sentiment_analysis}
\end{figure}

Figure \ref{fig:sentiment_analysis} illustrates the decision tree, offering more details on how the model predicts price changes. The tree shows that financial terms and ticker symbols are crucial for predicting upward or downward stock movements. For example, financial terms indicate price decreases, while terms related to market news and ticker symbols suggest price increases. Event-specific terms related to locations also play a role in the model’s predictions.

The findings suggest that mechanistic interpretability techniques like feature extraction via sparse autoencoders can enhance financial sentiment classification by reducing the reliance on costly fine-tuning while maintaining accuracy and generalizability. This approach enables efficient, interpretable sentiment classification in financial applications, supporting real-time sentiment tracking, multi-factor financial analysis, and regulatory compliance. Future work can extend this methodology to broader financial modeling tasks, improving the robustness and explainability of AI-driven decision-making in dynamic market environments.

A similar approach can be applied to other classification problems, such as developing a sentiment analysis by identifying Sparse Autoencoder (SAE) features in financial news associated with sentiment being positive, negative or neutral. 

\subsubsection{Bias Detection and Reduction}

In this experiment, we address bias reduction in the loan approval process by analyzing feature activations in sparse autoencoders (SAE). The approach begins by identifying potentially biased statements related to factors such as gender, marital status, and ethnicity in loan applications. A few hypothetical bias-related sentences were created for this purpose to simulate realistic biases observed in financial decision-making contexts. By analyzing feature activations in response to these statements, we identify features that consistently exhibit high activation, which may indicate biased model representations. This process helps pinpoint aspects of the model that could contribute to unfair or discriminatory outcomes.\footnote{The sentences and phrases used to identify bias-related features are purely for representational purposes and do not reflect the views of the author or their affiliations.}

Once potential bias-related features are identified, a system is developed to flag loan application text that contains biased language. When new loan application text is analyzed, the system checks the activation levels of the identified features. If these levels exceed a predefined threshold, the text is flagged for further review. For example, phrases like ``single mother with irregular income'' or ``high default rates among female entrepreneurs'' may trigger the bias alert. The goal of this detection system is to highlight language that may perpetuate bias, allowing for further scrutiny and reducing the risk of discriminatory decision-making.

The table below (Table~\ref{tab:loan_table}) presents loan applications that have been flagged or approved, along with their corresponding \textbf{bias ratio} and \textbf{loan text}. The \textbf{bias ratio} is defined as the percentage of bias-related features appearing among all activated features in the output. Bias ratio indicates the likelihood of biased language in the loan text, with higher values suggesting more significant bias.

\begin{table}[h]
\centering
\resizebox{\textwidth}{!}{
\begin{tabular}{|l|l|l|}
\hline
\textbf{status} & \textbf{bias\_ratio} & \textbf{loan\_text} \\ \hline
flagged          & 0.405289            & Applicant is a single mother with irregular income, so high-risk loan. \\ \hline
approved         & 0.283334            & Applicant has a stable job and good credit score, approval recommended. \\ \hline
flagged          & 0.363886            & Loan denied due to high default rates among female entrepreneurs. \\ \hline
flagged          & 0.395190            & Male applicants in this age group are more responsible with repayments compared to females. \\ \hline
approved         & 0.249879            & Application assessed based on income, and debt-to-income ratio. \\ \hline
\end{tabular}
}
\caption{Loan application status with corresponding bias ratio and loan text}
\label{tab:loan_table}
\end{table}

From the table, we observe that certain loan applications, such as \textit{Applicant is a single mother with irregular income, so high-risk loan} (flagged, bias ratio: 0.405289) and \textit{Loan denied due to high default rates among female entrepreneurs} (flagged, bias ratio: 0.363886), are flagged due to biased language tied to gender or income factors. These sentences exhibit higher bias ratios, indicating a higher likelihood of biased language. On the other hand, applications like \textit{Application assessed based on income, and debt-to-income ratio} (approved, bias ratio: 0.249879) do not contain biased language and are approved.

\subsubsection{Hallucination Reduction}

In this experiment, we aimed to enhance the reliability of the language model's output by mitigating potential hallucinations, specifically in the context of financial information. We utilized a feature selection method based on the activation magnitudes of various SAE features within the Google Gemma-2B model. By analyzing the features' activation levels, we selected those most relevant to the financial domain to help guide the hallucination reduction process. In addition to reviewing activation magnitudes, we also reviewed the labels assigned to the features to ensure that the selected features were meaningfully connected to financial concepts and terminology.

Table~\ref{tab:feature_table} shows the top features, including their mean activations. Features like \textbf{15857} (financial terms and information) and \textbf{15354} (financial news related to the stock market) were selected for their strong connection to the financial domain. These features are useful in distinguishing between accurate financial insights and potential hallucinations. The methodology involves calculating the percentage of activation for selected features relative to the total activation across all features for the given output. If the computed percentage falls below a predefined threshold, an additional retrieval step is triggered.

Specifically, retrieval-augmented generation (RAG) is used, a method that retrieves semantically relevant information from external sources to supplement the model's internal knowledge and improve factual grounding. For this experiment, a vector database containing processed SEC filings and financial documents was prepared; embeddings of document chunks were stored to allow efficient semantic retrieval when needed. When the percentage of finance-related feature activations was low, the model retrieved supporting information from this database to enhance output accuracy. A threshold of 20\% (chosen for illustration, and adjustable based on application needs) was set for triggering RAG, based on the sum of activations for the two selected features compared to the total activation across all features.

\begin{table}[h]
\centering
\footnotesize % Make the table even smaller
\begin{tabular}{|c|c|p{7cm}|} % Two columns with equal width
\hline
\textbf{Feature} & \textbf{Mean Activation} & \textbf{Feature Label} \\ \hline
\textbf{15857} & 10.842405 & \textbf{phrases related to financial terms and information} \\ \hline
5405  & 7.893589  & phrases related to market statistics and analysis \\ \hline
11912 & 2.604609  & text mentioning technical details and updates, including dates, times, and versions \\ \hline
6513  & 2.041639  & references to education and business institutions, particularly relating to specific programs, schools, and disciplines \\ \hline
14291 & 1.862908  & financial news related to the stock market \\ \hline
7906  & 1.781035  & mentions regarding customer interaction and feedback in a business context \\ \hline
\textbf{15354} & 1.751971  & \textbf{terms related to financial news and companies} \\ \hline
1213  & 1.395549  & dialogue and text indicating interactions among users in an online forum setting \\ \hline
8312  & 1.293682  & technical terms and administrative terms related to website administration and online forums \\ \hline
3553  & 1.130796  & names of places or locations. \\ \hline
544   & 1.129925  & terms related to medical conditions and medications, specifically focusing on cholesterol-lowering drugs \\ \hline
354   & 1.013154  & words related to financial markets, funds, and investments \\ \hline
6510  & 0.936306  & character sequences indicating the start of new sections or categories \\ \hline
15173 & 0.901733  & terms related to real estate and local government activities \\ \hline
3586  & 0.857892  & attends from the beginning-of-sequence marker to the first token in the sentence. \\ \hline
\end{tabular}
\caption{Top features selected for the financial domain, their mean activations}
\label{tab:feature_table}
\end{table}

To demonstrate the impact of the retrieval-augmented generation process, we pose the question "\textit{What are the current SEC requirements for Form 8-K filing deadlines?}"

Table~\ref{tab:rag_vs_no_rag} compares the model's responses without RAG and with RAG. 

The RAG output provides specific regulatory details and explicitly cites the source document: SEC's Final Rule 33-8400, "Additional Form 8-K Disclosure Requirements and Acceleration of Filing Date" (\href{https://www.sec.gov/rules/final/33-8400.htm}{17 CFR Parts 228, 229, 230, 239, 240 and 249}). The Without RAG output offers only general information about Form 8-K filings without any source citation. The activation threshold-based RAG process helps ensure that the model produces more accurate, factually consistent, and properly referenced responses, especially in critical financial regulatory contexts.

\begin{table}[h]
\centering
\footnotesize % Make the table even smaller
\begin{tabular}{|p{0.45\textwidth}|p{0.45\textwidth}|} % Two columns with equal width
\hline
\textbf{Without RAG} & \textbf{With RAG} \\
\hline
Companies must file Form 8-K to report significant corporate events in a timely manner, typically within four business days of the event. This keeps investors informed about material changes. & 
\textbf{Activation is below threshold. Triggering RAG.}

According to SEC regulations, Form 8-K must be filed within four business days after the occurrence of a triggering event for most items. However, Item 7.01 (Regulation FD Disclosure) requires filing on the same day. Item 4.02 (Non-Reliance on Previously Issued Financial Statements) has specific timing requirements based on when the determination is made. Late filings can result in SEC enforcement actions and loss of Form S-3 eligibility. Companies must maintain disclosure controls to ensure timely 8-K filings. Electronic filing through the EDGAR system is mandatory.

Source: SEC Final Rule 33-8400 (\href{https://www.sec.gov/rules/final/33-8400.htm}{17 CFR Parts 228, 229, 230, 239, 240 and 249}) \\
\hline
\end{tabular}
\caption{Comparison of model output with and without RAG}
\label{tab:rag_vs_no_rag}
\end{table}

This demonstrates the advantage of the feature-based selection with RAG approach in enhancing the factual accuracy and reliability of AI-generated financial insights.

Our proposed feature-driven conditional RAG approach, while still in its early stages, demonstrates the potential for improving the reliability of AI-generated financial insights. By monitoring the activation magnitudes of specific SAE features, we can identify situations where the model might be prone to generating speculative or unverifiable information. Triggering RAG in these cases helps ensure that the system incorporates factual, externally sourced data, ultimately enhancing the accuracy and trustworthiness of the generated output.

\subsection{Feature Steering}
\label{sec:feature steering}

As discussed in Section~\ref{sec:steering_sparse_autoencoders}, SAE’s feature steering is remarkably effective at modifying model outputs in specific, interpretable ways. It can be used to modify the model’s demeanor, preferences, stated goals, and biases; to induce it to make specific errors. Below we look at some examples.

\subsubsection{ Modifying Credit Risk Related Features}

In this experiment, we use feature steering to modify a model’s sentiment analysis of financial statements by shifting its focus toward credit risk. Instead of relying solely on prompt engineering, we steer specific features to adjust the model’s interpretation, demonstrating how steering can enhance context-aware financial analysis without requiring fine-tuning. This experiment was conducted using Neuronpedia, a DeepMind-supported platform referred to earlier.

\begin{figure}[H]
    \centering
    \includegraphics[width=1\textwidth]{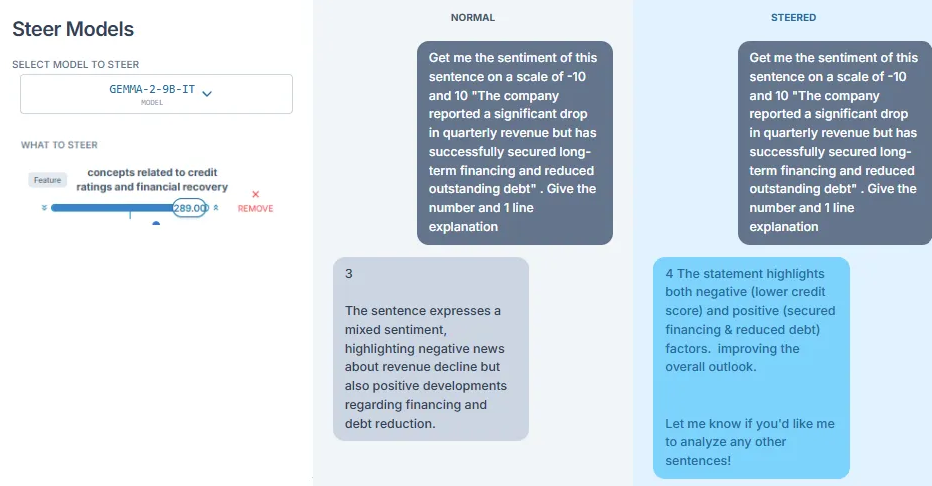}
    \caption{Feature steering influence on sentiment analysis in financial statements}
    \label{fig:feature_steering_credit_risk}
\end{figure}

Figure \ref{fig:feature_steering_credit_risk} demonstrates how feature steering influences sentiment interpretation in financial analysis. Initially, without steering, the model assigns a \textbf{mixed sentiment (3)} to the sentence: \textit{The company reported a significant drop in quarterly revenue but has successfully secured long-term financing and reduced outstanding debt}. The model primarily focuses on the negative impact of the \textit{revenue decline}, interpreting it as a sign of financial instability, while only partially considering the positive aspects of secured financing and debt reduction.

With feature steering applied at a high magnitude to the feature \textit{"concepts related to credit ratings and financial recovery"}, the sentiment shifts to \textbf{4}, highlighting both the \textit{negative (lower credit score)} and \textit{positive (secured financing \& debt reduction)} aspects, ultimately improving the overall outlook. This change occurs because steering shifts the model’s attention toward financially stabilizing factors, ensuring a more balanced and domain-specific interpretation.

This process might eliminate the need for extensive prompt engineering, as the model automatically emphasizes credit recovery factors without requiring manually reworded inputs like \textit{Analyze from a credit risk perspective}. Additionally, steering serves as a cost-effective alternative to fine-tuning, allowing existing models to dynamically adjust without requiring thousands of labeled financial statements.

Another key question is 'How much should we steer?' Feature steering alters model behavior in different evaluations and can introduce unintended trade-offs. Understanding whether its effects are predictable from feature activations or broader and harder to anticipate is crucial to refine its application (refer \textit{Toward Monosemanticity: Decomposing Language Models }~\cite{Anthropic2024}). Some studies have shown that excessive feature steering can lead to entangled representations.

\subsubsection{Credit Risk-Focused Steering on Sentiments}

A common financial application of AI models is extracting financial sentiment from a set of unstructured data, documents, or news sources. Basically, it is a classification problem of converting text input into a few classification categories.  

In our example, we have collected a set of manually-annotated sentences with the following 5 categories: very positive, somewhat positive, neutral, somewhat negative, and negative. These categories have been human labeled according to a severity of a news statement as it relates to the financial health of a company or the company's ability to sustain debt repayment. Using the Gemma-2B model with the following prompt: \emph{ Can you rate  the following sentence as a sentiment, very positive, somewhat positive, neutral, somewhat negative, very negative}, we are able to compare the unsteered and steered version of the sentiment classification.  The experiment makes an assumption that steering a particular feature related to Credit Risk will improve the quality of classification using LLMs.

\begin{figure}[h]
    \centering
    \includegraphics[width=\textwidth]{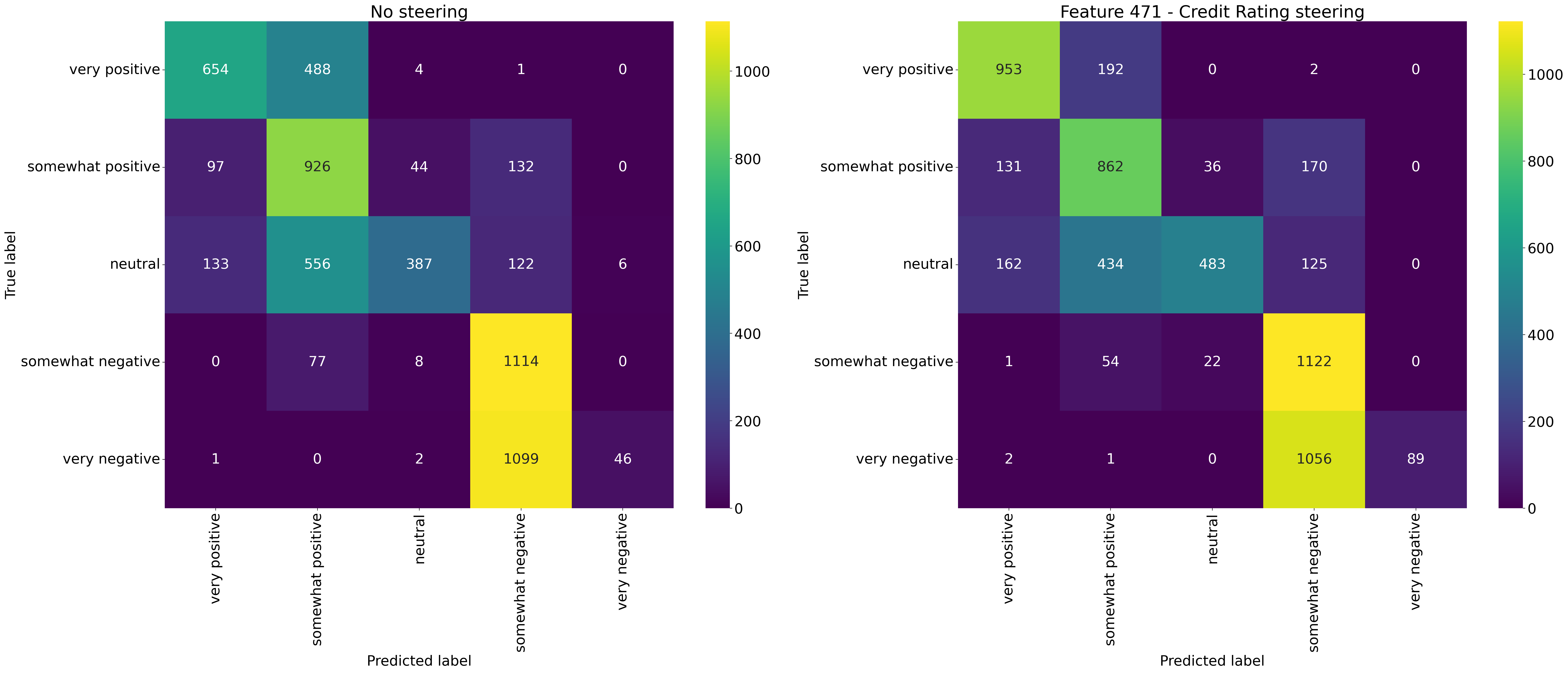}
    \caption{Confusion matrices comparing sentiment classification performance before (left) and after (right) applying credit risk-focused steering}
    \label{fig:cfm}
\end{figure}

The following confusion matrices (Figure~\ref{fig:cfm}) compare sentiment classification results before and after applying steering for feature 471, listed in Table~\ref{tab:feature_description}, which is identified as related to "credit risk". The left matrix represents classification results without steering, while the right matrix shows improvements after steering. 

Without steering, the model frequently misclassified ``very positive'' and ``somewhat positive'' sentiments, with 769 instances of ``very positive'' misclassified as ``somewhat positive.'' Similarly, 132 neutral labels were mistakenly assigned to ``somewhat positive.'' Misclassification also occurred between ``somewhat negative'' and ``very negative,'' with 46 instances of ``very negative'' incorrectly predicted as ``somewhat negative.''

After applying credit risk-focused steering, classification accuracy improved across all sentiment categories. The correct classification of ``very positive'' increased from 654 to 953 (+299), while misclassifications into ``somewhat positive'' dropped by 296. Neutral misclassification errors declined from 132 to 36, reducing confusion with positive sentiments. The distinction between ``somewhat negative'' and ``very negative'' was also refined, reducing false negatives and improving overall credit risk sentiment differentiation.

Overall, steering significantly reduced false positives and negatives, aligning sentiment classification with credit risk considerations. The refined predictions demonstrate the effectiveness of domain-specific steering in financial sentiment analysis.

\section{Challenges}

Mechanistic interpretability in finance faces key challenges related to scalability, generalization, and feature-level transparency. While methods such as logit attribution, activation patching, and sparse autoencoder-based steering have improved AI explainability, scalability remains a major concern, especially as financial models grow to GPT-4 scale and beyond. Manual circuit analysis is infeasible for large models, and automated interpretability tools are still in early development. Furthermore, many existing techniques fail to generalize effectively to real-world financial applications, as models often rely on abstract representations that do not directly map to human-understandable financial concepts. The lack of ground-truth benchmarks for financial interpretability further complicates evaluating and validating AI-driven decision-making \citep{elhage2022b}.

Another critical challenge is the evolving nature of financial markets and regulations. AI models trained on historical data may struggle to adapt to changing risk environments, new financial instruments, and updated compliance laws. Dynamic interpretability pipelines that allow for real-time steering and adaptive model explanations are necessary to ensure AI models remain aligned with financial industry standards. The reliance on human evaluations in interpretability research also presents risks, as subjective analyses can introduce inconsistencies when explaining financial model decisions \citep{Rauker2023}.

To address these challenges, fully integrated interpretability tooling must combine techniques from various research efforts, such as Distill-style visualizations, Redwood’s tools, and Nanda’s TransformerLens \citep{nanda2022}. A unified framework could streamline analysis workflows, making it easier for financial professionals to debug models, trace decisions, and refine AI-driven predictions. Mechanistic interpretability is particularly valuable for financial institutions looking to improve credit risk modeling, algorithmic trading transparency, fraud detection, and regulatory compliance. By enhancing AI explainability in finance, institutions can build trust, improve governance, and ensure models remain robust in dynamic economic conditions \citep{hendrycks2023}.

\section*{Acknowledgement}

We are grateful to Jeffrey Nisen, Zirui Ding, Oksana Kityachik, Andrew Whiteley, and Navneet Mathur - our colleagues at Barclays for their comments, corrections and inspiration.

\section{Conclusion}

Mechanistic interpretability offers a promising route to dissect and guide AI-driven financial decisions with greater transparency and control. By combining observation-based methods (like logit attribution and sparse autoencoders) with intervention-based techniques (such as activation patching and feature steering), financial practitioners can pinpoint the neurons and attention heads most influential in tasks such as fraud detection, credit scoring, and compliance. Our experiments showed that these interpretability tools not only help isolate crucial risk or sentiment indicators but also allow dynamic model adjustments—cutting down on costly fine-tuning and enabling more precise responses in high-stakes scenarios.

In practice, mechanistic interpretability enhances AI reliability in finance. By replacing extensive prompt engineering and reducing reliance on fine-tuning, steering provides a cost-effective alternative for refining financial AI applications dynamically. These insights reinforce the need for continued research in the field of \textbf{financial AI safety}, particularly in the alignment of LLMs with regulatory standards and risk governance frameworks.

Despite its benefits, this approach faces notable challenges. Large-scale financial models can exceed current methods’ capacity to manually analyze circuits, and automated interpretability frameworks remain a work in progress. Financial features often blend abstract economic factors that are difficult to map to human-understandable concepts, and evolving markets or regulatory changes can outpace models’ training data. Additionally, identifying a particular neuron or attention head does not always clarify the exact influence of features such as credit score or volatility on the final decision—requiring integration with traditional model-agnostic methods for a truly comprehensive explanation.

\subsection{Future Work}

Looking ahead, the adoption of integrated interpretability pipelines will be crucial for ensuring AI-driven finance remains both transparent and adaptive. Combining mechanistic interpretability with feature-level attributions and real-time intervention tools can mitigate risks stemming from bias or market shifts. Future work can focus on integrating mechanistic interpretability with \textbf{feature-level attributions}, ensuring that AI decisions remain explainable not just at the neuron level but also in terms of real-world financial indicators. One promising direction is to integrate traditional statistical modeling techniques that utilize financial time series data with unstructured data processed by LLMs. This hybrid approach offers an explainable framework for combining structured and unstructured data, unlocking access to the vast potential of unstructured information.

As these methods mature, they will pave the way for robust, interpretable financial AI applications that can be safely aligned with regulatory standards and responsive to the dynamic demands of modern finance.

% \bibliography{main}
% \bibliographystyle{tmlr}

\input{main.bbl}
\end{document}

%% file: math_commands.tex
\usepackage{amsmath,amsfonts,bm}

\def\eqref#1{equation~\ref{#1}}
\def\1{\bm{1}}

\DeclareMathAlphabet{\mathsfit}{\encodingdefault}{\sfdefault}{m}{sl}
\SetMathAlphabet{\mathsfit}{bold}{\encodingdefault}{\sfdefault}{bx}{n}